\documentclass[pra,aps,amsmath,amssymb,showkeys,showpacs,twocolumn]{revtex4}
\newcommand{\hh}{\mathcal{H}}
\newcommand{\pen}{\openone}
\newcommand{\lnp}{\mathcal{L}}
\newcommand{\lsp}{\mathcal{L}_{+}}
\newcommand{\am}{{\mathsf{A}}}
\newcommand{\ax}{{\mathsf{X}}}

\newcommand{\cla}{{\mathcal{A}}}
\newcommand{\clb}{\mathcal{B}}
\newcommand{\cli}{{\mathcal{I}}}

\newcommand{\cmb}{\mathbb{B}}
\newcommand{\bro}{\boldsymbol{\rho}}
\newcommand{\bdl}{\boldsymbol{\delta}}
\newcommand{\bmg}{\boldsymbol{\omega}}
\newcommand{\tr}{\mathrm{tr}}
\newcommand{\ron}{{\mathrm{ran}}}
\newcommand{\dig}{\mathrm{diag}}
\newcommand{\rg}{\mathrm{g}}
\begin{document}
\clearpage
\preprint{}

\title{Uncertainty relations for quantum coherence with respect to mutually unbiased bases}

\author{Alexey E. Rastegin}
\affiliation{Department of Theoretical Physics, Irkutsk State University, Russia}

\begin{abstract}
The concept of quantum coherence, including various ways to quantify the degree of coherence
with respect to the prescribed basis, is currently the subject of
active research. The complementarity of quantum coherence in different
bases was studied by deriving upper bounds on the sum of
the corresponding measures. To obtain a two-sided estimate, lower
bounds on the coherence quantifiers are also of interest. Such
bounds are naturally referred to as uncertainty relations for
quantum coherence. We obtain new uncertainty relations for
coherence quantifiers averaged with respect to a set of mutually
unbiased bases (MUBs). To quantify the degree of coherence, the relative
entropy of coherence and the geometric coherence are used. Further, we also derive novel state-independent uncertainty relations
for a set of MUBs in terms of the min-entropy.
\end{abstract}



\maketitle

\pagenumbering{arabic}
\setcounter{page}{1}

\section{Introduction}\label{sec1}

The principle of state superposition is one of the cornerstones of
quantum theory. It is closely related to the problem of properly
quantifying coherence at the quantum level, which has attracted 
considerable attention \cite{bcp14,plenio16}. This line of study
is part of more general efforts to understand the strengths and
limitations of nonclassical correlations \cite{adesso16jpa}. In
principle, a given quantum state can be represented with respect
to an arbitrary basis. However, only a few bases may actually be
preferred for fundamental physical reasons. When we deal with the
measurement problem, the existence of some preferable basis is one
of the principal questions to be resolved \cite{zurek81}. Studies of
the thermodynamic properties of small systems at low temperatures also
imply the use of a quite concrete representation of possible
mixed states \cite{horodecki15,rudolx15,ngour15}. Analysis of
protocols and algorithms for quantum information processing
assumes that transformations of quantum carriers will be related
to the prescribed orthonormal bases. Thus, the role of quantum
coherence should be understood in order to realize efficient quantum
computation. Results of recent studies
\cite{hillery16,hfan2016} have supported this opinion. Duality
relations between the coherence and path information were examined
in \cite{bera15,bagan16}.

Some complementarity relations for quantum coherence were studied
in \cite{hall15}. They claimed upper bounds on the sum of
coherence quantifiers taken with respect to mutually unbiased
bases (MUBs). Uncertainty bounds for coherence can be obtained using
previously obtained uncertainty relations of the usual form. Even in
the case of conjugate variables, uncertainty relations still
attract attention \cite{werner16,cancon16,gegen17}. Basic
developments within the entropic approach to quantum uncertainty
are reviewed in \cite{ww10,brud11,cbtw15}. The famous
Maassen--Uffink result \cite{maass} has been applied in many
problems including entropic relations in the presence of quantum
memory \cite{BCCRR10}. These results were used in deriving
uncertainty relations for quantum coherence
\cite{pati16,pzflf16,baietal6}. In this work, we obtain new
uncertainty bounds for the averaged quantum coherence taken with
respect to MUBs. The relevant entropic uncertainty relations
will be mentioned in appropriate places in the
text. The novelty of our results is threefold. First, lower bounds
on coherence measures only for MUBs had not been 
addressed separately. Second, we obtain some uncertainty relations for the
averaged geometric coherence. Third, we present new
state-independent uncertainty relations for a set of MUBs in terms
of the averaged min-entropy.

\section{Preliminaries}\label{sec2}

Let $\lnp(\hh)$ be the space of linear operators on a
$d$-dimensional Hilbert space $\hh$. The set of positive
semidefinite operators will be denoted by $\lsp(\hh)$. By
$\ron(\ax)$, we mean the range of operator $\ax$. For each
$\ax\in\lnp(\hh)$, we define $|\ax|\in\lsp(\hh)$ as the unique
square root of $\ax^{\dagger}\ax$. The eigenvalues of $|\ax|$
counted with multiplicities are the singular values
$\sigma_{i}(\ax)$ of $\ax$. We will further refer to the following
two norms,
\begin{align}
\|\ax\|_{1}&:=\sum_{i=1}^{d} \sigma_{i}(\ax)
\, , \label{nrm1}\\
\|\ax\|_{\infty}&:=\max\bigl\{\sigma_{i}(\ax):{\>}1\leq{i}\leq{d}\bigr\}
\, , \label{nrmin}
\end{align}
which are known as the trace and spectral norms, respectively. The
state of a quantum system is described by the density matrix
$\bro\in\lsp(\hh)$ normalized as $\tr(\bro)=1$. Then the von
Neumann entropy of $\bro$ is defined as
\begin{equation}
S_{1}(\bro):=\!{}-\tr(\bro\ln\bro)
\, . \label{vnent}
\end{equation}

Let us consider orthonormal bases
$\clb=\bigl\{|b_{i}\rangle\bigr\}$ and
$\clb^{\prime}=\bigl\{|b_{j}^{\prime}\rangle\bigr\}$ in the
$d$-dimensional space $\hh$. They are said to be mutually unbiased
if and only if for all $i$ and $j$,
\begin{equation}
\bigl|\langle{b}_{i}|b_{j}^{\prime}\rangle\bigr|=\frac{1}{\sqrt{d}}
\ . \label{twbs}
\end{equation}
Several orthonormal bases form a set of MUBs when each pair in them is mutually unbiased. MUBs are applied in many quantum
information problems. They also form a structure with interesting
properties and links to discrete mathematics. As even a listing of
all such topics may be incomplete, we refer to the literature
(see \cite{bz10} and references therein). In general, the maximal
number of MUBs in $d$ dimensions is an open problem. For a prime
power $d$, we can certainly construct $d+1$ MUBs.

A rigorous framework for the quantification of coherence was
developed in \cite{bcp14}. To the given orthonormal basis
$\clb=\bigl\{|b_{i}\rangle\bigr\}$, we assign the set of all density
matrices that are diagonal in this basis,
\begin{equation}
\bdl=\sum_{i=1}^{d} \delta_{i}\,|b_{i}\rangle\langle{b}_{i}|
\, . \label{incod}
\end{equation}
Density matrices of the form (\ref{incod}) constitute the set
$\cli_{\clb}$ of states incoherent with respect to $\clb$. Keeping
in mind the chosen basis, we ask how far the given state is from
those states that are completely incoherent in this basis. The
authors of \cite{bcp14} listed general conditions for quantifiers
of coherence. Additional conditions imposed on coherence measures
were considered in \cite{plenio16}. The imposed conditions allow
us to identify classes of functionals that can be used as proper
coherence measures. In this paper, we will use the relative
entropy of coherence and the geometric coherence.

The concept of quantum relative entropy, or divergence, is basic in
quantum information theory \cite{nielsen,vedral02}. For density
matrices $\bro$ and $\bmg$, this quantity is expressed as
\cite{nielsen}
\begin{equation}
D_{1}(\bro||\bmg):=
\begin{cases}
\tr(\bro\ln\bro\,-\bro\ln\bmg) \,,
& \text{if $\ron(\bro)\subseteq\ron(\bmg)$} \, , \\
+\infty\, , & \text{otherwise} \, .
\end{cases}
\label{relan}
\end{equation}
Although the relative entropy cannot be treated as a metric, it
provides a very natural measure of distinguishability of quantum
states. Following \cite{bcp14}, we define the coherence measure
\begin{equation}
C_{1}(\clb|\bro):=
\underset{\bdl\in\cli_{\clb}}{\min}\,D_{1}(\bro||\bdl)
\, . \label{c1df}
\end{equation}
The minimization is easy and results in the formula \cite{bcp14}
\begin{equation}
C_{1}(\clb|\bro)=S_{1}(\bro_{\dig})-S_{1}(\bro)
\, , \label{c1for}
\end{equation}
where the diagonal state
\begin{equation}
\bro_{\dig}:=\dig\bigl(
\langle{b}_{1}|\bro|b_{1}\rangle,\ldots,\langle{b}_{d}|\bro|b_{d}\rangle
\bigr)
\, . \label{rhodig}
\end{equation}
We can represent $S(\bro_{\dig})$ as the Shannon entropy
calculated with probabilities
$p_{i}(\clb|\bro)=\langle{b}_{i}|\bro|b_{i}\rangle$, namely,
\begin{equation}
S_{1}(\bro_{\dig})=H_{1}(\clb|\bro):=-\sum_{i=1}^{d} p_{i}(\clb|\bro)\, \ln{p}_{i}(\clb|\bro)
\, . \label{shlin}
\end{equation}
For the general properties of (\ref{c1df}), see the relevant sections in
\cite{bcp14,plenio16}. It seems that the relative entropy of
coherence is the most justifiable measure. Together with
(\ref{relan}), other quantum divergences were considered,
including the quasi-entropies of Petz \cite{petz86}. It is for this
reason that we designate the considered entropic quantities by the
subscript $1$. Coherence quantifiers induced by quantum
divergences of the Tsallis type were addressed in
\cite{rastpra16}. It was shown that such quantifiers do not
support a simple form similar to (\ref{c1for}). Coherence
monotones based on R\'{e}nyi divergences were considered in
\cite{chitam2016,shao16,skwgb16}.

There are also several distance-based quantifiers of coherence
\cite{bcp14,plenio16}. We will use the geometric coherence, which
was introduced in terms of the quantum fidelity
\cite{uhlmann76,jozsa94}. The fidelity of density matrices $\bro$
and $\bmg$ is expressed as
\begin{equation}
F(\bro,\bmg)=\bigl\|\sqrt{\bro}\,\sqrt{\bmg}\bigr\|_{1}^{2}
\, . \label{fiddf}
\end{equation}
This definition follows Jozsa \cite{jozsa94}. Another way is to
set the fidelity as the square root of (\ref{fiddf})
\cite{nielsen}. The fidelity ranges between $0$ and $1$, taking the
value $1$ for two identical states. Hence, the difference
$1-F(\bro,\bmg)$ can be treated as a distance measure. Strictly
speaking, a legitimate metric is obtained after the
root of this difference is extracted and is usually called the sine distance
\cite{gln2005,rastsin}. Following the general approach, the
geometric coherence of $\bro$ with respect to the basis $\clb$ is
defined as \cite{plenio16}
\begin{equation}
C_{\rg}(\clb|\bro):=1-
\underset{\bdl\in\cli_{\clb}}{\max}\,F(\bro,\bdl)
\, . \label{cgdf}
\end{equation}
For the properties of this coherence quantifier, see subsection
III.C.3 of \cite{plenio16}. For a pure state, the geometric
coherence has a very convenient form:
\begin{equation}
C_{\rg}(\clb|\psi)=1-\underset{i}{\max}\,\bigl|\langle{b}_{i}|\psi\rangle\bigr|^{2}
\, . \label{cgpor}
\end{equation}
For impure states, we have only a two-sided estimate on
the geometric coherence. For the given density matrix $\bro$ and
orthonormal basis $\clb$, the index of coincidence is introduced
as
\begin{equation}
J(\clb|\bro):=\sum_{i=1}^{d} p_{i}(\clb|\bro)^{2}
\ , \label{icdf}
\end{equation}
where $p_{i}(\clb|\bro)=\langle{b}_{i}|\bro|b_{i}\rangle$.
The authors of \cite{geomes17} have proved that
\begin{align}
&\frac{d-1}{d}\,
\biggl\{1-
\sqrt{\,1+\frac{d}{d-1}\,\bigl[J(\clb|\bro)-\tr(\bro^{2})\bigr]}
\,\biggr\}
\nonumber\\
&\leq{C}_{\rg}(\clb|\bro)\leq1-\underset{i}{\max}\>p_{i}(\clb|\bro)
\, . \label{twogeos}
\end{align}
This two-sided estimate is based on the concept of sub- and
super-fidelities proposed in \cite{uhlmann09}. It seems that the
quantifier (\ref{cgdf}) deserves to be studied more widely. For
this reason, we will formulate uncertainty relations in terms of
the geometric coherence.

Additionally, we recall some results concerning special forms of
the uncertainty relations. Two of our relations are based on the
following inequality derived for $M$ MUBs in \cite{molm09}:
\begin{equation}
\sum_{t=1}^{M} J(\clb_{t}|\bro)
\leq\tr(\bro^{2})+\frac{M-1}{d}
\leq1+\frac{M-1}{d}
\ . \label{indk}
\end{equation}
Assuming the existence of $d+1$ MUBs, one can show that 
inequality (\ref{indk}) is saturated here. The authors of
\cite{molm09} further used (\ref{indk}) to obtain uncertainty
relations in terms of the Shannon entropy. In 
\cite{rastmub}, we extended this approach to the R\'{e}nyi and Tsallis
entropies.

Let us recall also one result of \cite{imai07}. Any
quantum measurement can be described by the set $\cla$ of elements
$\am_{j}\in\lsp(\hh)$ such that the completeness relation holds:
\begin{equation}
\sum\nolimits_{j} \am_{j}=\pen
\, . \label{cmrl}
\end{equation}
The set $\cla$ is a positive operator-valued measure (POVM)
\cite{nielsen}. We do not specify the range of summation in
(\ref{cmrl}), as the number of different outcomes in a POVM
measurement can exceed the dimensionality $d$. The trace
$\tr(\am_{j}\bro)$ gives the probability of the $j$-th outcome. Let
$\bigl\{\cla_{1},\ldots,\cla_{M}\bigr\}$ be a set of $M$ POVMs,
and let some index $j(t)$ be assigned to each $t=1,\ldots,M$. For
an arbitrary state $\bro$, it holds that \cite{imai07}
\begin{equation}
\sum_{t=1}^{M} p_{j(t)}(\cla_{t}|\bro)
\leq
1+\biggl(\,\sum_{s\neq{t}}{\Bigl\|\sqrt{\am_{j(s)}^{(s)}}\sqrt{\am_{j(t)}^{(t)}}\Bigr\|_{\infty}^{2}}\biggr)^{\!1/2}
. \label{mim6}
\end{equation}
This upper bound leads to uncertainty relations of the
Landau--Pollak type for more than two measurements.
Indeed, we can choose the indices $j(t)$ so that the left-hand side of
(\ref{mim6}) will include the maximal probability for each
measurement. Formula (\ref{mim6}) provides a nontrivial bound
for $M\geq2$. Further studies of the Landau--Pollak uncertainty
relations for POVM measurements were reported in \cite{bosyk14}.

\section{Main results}\label{sec3}

We are now ready to formulate the uncertainty relations for quantum
coherence taken with respect to a set of MUBs. Let us begin the
presentation with lower bounds on the averaged relative entropy of
coherence. The following statement is presented.

\newtheorem{prp1}{Proposition}
\begin{prp1}\label{pan1}
Let $\cmb=\bigl\{\clb_{1},\ldots,\clb_{M}\bigr\}$ be a set of MUBs
in the $d$-dimensional Hilbert space $\hh$. For any state $\bro$, the
averaged relative entropy of coherence obeys
\begin{align}
&\frac{1}{M}\,\sum_{\clb\in\cmb} C_{1}(\clb|\bro)\geq
\nonumber\\
&\ln\!\left(\frac{Md}{\tr(\bro^{2}){\,}d+M-1}\right)
-S_{1}(\bro)
\, . \label{bound1}
\end{align}
\end{prp1}

{\bf Proof.} According to (\ref{c1for}), the left-hand side of
(\ref{bound1}) is represented as
\begin{align}
&{}-S_{1}(\bro)+\frac{1}{M}\,\sum_{t=1}^{M} H_{1}(\clb_{t}|\bro)\geq
\nonumber\\
&{}-S_{1}(\bro)+\ln\!\left(\frac{Md}{\tr(\bro^{2}){\,}d+M-1}\right)
 . \label{prf1}
\end{align}
Here, we used the lower bound on the averaged Shannon entropy
proved in \cite{molm09,rastmub}. $\blacksquare$

The statement of Proposition \ref{pan1} provides a state-dependent
lower bound on the averaged relative entropy of coherence taken
with respect to a set of MUBs. For any pure state $|\psi\rangle$,
it reduces to
\begin{equation}
\frac{1}{M}\,\sum_{\clb\in\cmb} C_{1}(\clb|\psi)
\geq\ln\!\left(\frac{Md}{d+M-1}\right)
\, . \label{bound1p}
\end{equation}
If $d$ is a prime power, we certainly have $d+1$ MUBs. In this
case, the lower bound (\ref{bound1}) reads as
\begin{equation}
\frac{1}{d+1}\,\sum_{t=1}^{d+1} C_{1}(\clb_{t}|\bro)
\geq\ln\!\left(\frac{d+1}{\tr(\bro^{2})+1}\right)
-S_{1}(\bro)
\, . \label{bound1d}
\end{equation}
It is essential that arguments of the logarithm in (\ref{bound1})
and (\ref{bound1d}) depend on the purity $\tr(\bro^{2})$. For the
completely mixed state $\bro_{*}=\pen/d$, these bounds are
obviously saturated.

The authors of \cite{pati16,pzflf16} derived uncertainty relations
for the relative entropy of coherence taken with respect to
different bases. These results follow from the entropic uncertainty
relation obtained in \cite{lmfan15}. The method of
that paper was inspired in turn by the relative entropy approach to
entropic uncertainties proposed in \cite{ccyz12}. The authors of
\cite{baietal6} dealt mainly with relations for the relative
entropy of coherence in two measurement bases. In the qubit case,
they also considered the uncertainty bounds on the
$\ell_{1}$-norm of coherence. To justify the significance of
relations of the form (\ref{bound1}), we should compare them with
the results of \cite{pati16,pzflf16}. Let us begin with the case
of pure states. For the averaged coherence, the corresponding
inequality of \cite{pati16,pzflf16} can be written as
\begin{equation}
\frac{1}{M}\,\sum_{\clb\in\cmb} C_{1}(\clb|\psi)
\geq\!{}-\frac{\ln{m}(\cmb)}{M}
\, . \label{mound1p}
\end{equation}
The principal quantity $m(\cmb)$ should be found by solving a
certain problem of two-stage maximization \cite{lmfan15}. This
sufficiently difficult problem is essentially simplified for MUBs.
We then deal with a quantity obtained from the multiple sum with
$(M-2)$ indices, each of which runs $d$ different values. The
factor to be added is independent of these indices and equal to
$d^{1-M}$. Using (\ref{mound1p}) for a set of MUBs, we substitute
\begin{equation}
m(\cmb)=d^{M-2}\,d^{1-M}=\frac{1}{d}
\, . \label{mounm}
\end{equation}
Thus, the right-hand side of (\ref{mound1p}) becomes $\ln{d}/M$. The
latter may compete with (\ref{bound1p}) only when the number $M$
is sufficiently small compared with $d$. For a prime power
$d$, the maximal number of MUBs is certainly equal to $d+1$. For
$M=d+1$, the right-hand side of (\ref{bound1p}) is
$\ln(d+1)-\ln2$, whereas the right-hand side of (\ref{mound1p}) is
$\ln{d}/(d+1)$. To each $d$, we can assign some value $M_{1}$ such
that for all allowed $M\geq{M}_{1}$, the result (\ref{bound1p}) is
stronger than (\ref{mound1p}). For the given $M_{1}$, we
actually have the corresponding interval of dimensionality. Table \ref{tab1} lists such intervals for
$M_{1}=3,4,5,6,7$. Thus, our uncertainty relations for quantum
coherence are relevant, at least for pure states.

\begin{table}
\begin{center}
\caption{\label{tab1}Intervals of values of $d$ for which 
formula (\ref{bound1p}) gives a stronger bound.} \vskip0.1cm
\begin{tabular}{c|c|c|c|c}
\hline
$M_{1}=3\,$ & $M_{1}=4$ & $M_{1}=5$ & $M_{1}=6$ & $M_{1}=7$ \\
\hline
$2\div{20}\,$ & $\,21\div{243}\,$ & $\,244\div{3104}\,$ & $\,3105\div{46625}\,$ & $\,46626\div{823500}\,$ \\
\hline
\end{tabular}
\end{center}
\end{table}

Our second result concerns the uncertainty relations for the geometric
coherence taken with respect to a set of MUBs. Lower bounds on the
averaged geometric coherence of a pure state are posed as follows.

\newtheorem{prp2}[prp1]{Proposition}
\begin{prp2}\label{pan2}
Let $\cmb=\bigl\{\clb_{1},\ldots,\clb_{M}\bigr\}$ be a set of MUBs
in the $d$-dimensional Hilbert space $\hh$. For any state $\bro$,
 the average geometric coherence obeys
\begin{align}
&\frac{1}{M}\,\sum_{\clb\in\cmb} C_{\rg}(\clb|\bro)
\geq
\nonumber\\
&\frac{d-1}{d}-\frac{\sqrt{d-1}}{d\,\sqrt{M}}\,\sqrt{Md-1-(Md-d)\,\tr(\bro^{2})}
\ . \label{bound21m}
\end{align}
For a pure state $|\psi\rangle$, we also have
\begin{equation}
\frac{1}{M}\,\sum_{\clb\in\cmb} C_{\rg}(\clb|\psi)
\geq
1-\frac{1}{M}
\,\biggl(
1+\sqrt{\frac{M^{2}-M}{d}}
\>\biggr)
\, . \label{bound22}
\end{equation}
\end{prp2}

{\bf Proof.} The left-hand side of (\ref{twogeos}) is a convex and
decreasing function of $J(\clb|\bro)$. By Jensen's inequality,
we obtain
\begin{align}
&\frac{1}{M}\,\sum_{\clb\in\cmb} C_{\rg}(\clb|\bro)
\geq
\frac{d-1}{d}
\,\biggl\{1-{}
\biggr.
\nonumber\\
&\biggl.
\sqrt{\,1-\frac{\tr(\bro^{2})\,d}{d-1}+\frac{d}{(d-1)M}\,\sum\nolimits_{t=1}^{M}J(\clb_{t}|\bro)}
\,\biggr\}
\, . \label{geos}
\end{align}
Combining (\ref{indk}) with (\ref{geos}) finally gives (\ref{bound21m}).

Let us prove the claim (\ref{bound22}). It follows from (\ref{cgpor}) that
\begin{equation}
\frac{1}{M}\,\sum_{t=1}^{M} C_{\rg}(\clb_{t}|\psi)=
1-\frac{1}{M}\,\sum_{t=1}^{M} p_{\max}(\clb_{t}|\psi)
\, , \label{prf21}
\end{equation}
where
$p_{\max}(\clb|\psi):=\max\{p_{i}(\clb|\psi):{\>}1\leq{i}\leq{d}\}$.
The sum of the maximal probabilities on the right-hand side of
(\ref{prf21}) can be estimated by applying (\ref{mim6}) to the
case of MUBs. The right-hand side of (\ref{mim6}) takes a simple
form, so 
\begin{equation}
\sum_{t=1}^{M} p_{\max}(\clb_{t}|\psi)\leq
1+\sqrt{\frac{M^{2}-M}{d}}
\, . \label{prf25}
\end{equation}
Combining (\ref{prf21}) with (\ref{prf25}) completes the proof of
(\ref{bound22}). $\blacksquare$

The statement of Proposition \ref{pan2} gives lower bounds on the
geometric coherence averaged with respect to
several MUBs. For pure states, we have arrived at 
two different formulas. Indeed, the relation (\ref{bound21m}) then
reduces to
\begin{equation}
\frac{1}{M}\,\sum_{\clb\in\cmb} C_{\rg}(\clb|\psi)
\geq
\frac{d-1}{d}\left(1-\frac{1}{\sqrt{M}}\right)
 . \label{bound21}
\end{equation}
Inspection reveals that neither of the lower bounds [(\ref{bound22}) and (\ref{bound21})] should be
disregarded. For the prime power $d$, we can use up to $d+1$ MUBs. It
can be checked that
\begin{equation}
\frac{1}{d}\left(1+\frac{d-1}{\sqrt{d+1}}\right)<
\frac{1}{d+1}+\frac{1}{\sqrt{d+1}}
\ . \nonumber
\end{equation}
Here, the lower bound (\ref{bound21}) is stronger. The same fact
is obvious for $M=d$, when $d$ MUBs exist in
$d$ dimensions. For sufficiently small $M$, however, the lower
bound (\ref{bound22}) is better than (\ref{bound21}). Such
behavior was already mentioned in the context of the separability
conditions in terms of the maximal probabilities \cite{rastsep}. On
the other hand, the distinctions between the lower bounds (\ref{bound22})
and (\ref{bound21}) are relatively small for all
$M$ that could be allowed here.

At this stage, we ask whether inequality (\ref{prf25}) can
lead to some analog of (\ref{bound1}). The answer is positive in
the sense that state-independent lower bounds on the averaged
relative entropy of coherence actually follow from (\ref{prf25}).
To obtain such bounds, we use convexity and the decrease of the
function $x\mapsto\!{}-\ln{x}$, together with
$H_{1}(\clb|\bro)\geq-\ln{J}(\clb|\bro)$ and
$J(\clb|\bro)\leq{p}_{\max}(\clb|\bro)$. Hence, we obtain
\begin{align}
\frac{1}{M}\,\sum_{t=1}^{M} H_{1}(\clb_{t}|\bro)
&\geq\!{}-\ln\!\left(\frac{1}{M}\,\sum\nolimits_{t} J_{t}\right)
\nonumber\\
&\geq\ln\biggl(\frac{M\sqrt{d}}{\sqrt{d}+\sqrt{M^{2}-M}}\biggr)
\, . \label{resobs}
\end{align}
Simple calculations show that the right-hand side of (\ref{resobs}) is
less than the right-hand side of (\ref{bound1p}). This result is 
evidence that inequality (\ref{prf25}) is especially useful in
estimates directly expressed via the maximal probabilities. To
illustrate this conclusion, we will obtain new uncertainty
relations for MUBs in terms of the averaged min-entropy.

By taking the maximum among the probabilities $p_{i}(\clb|\bro)$, the
min-entropy is defined as
\begin{equation}
H_{\infty}(\clb|\bro):=\!{}-\ln{p}_{\max}(\clb|\bro)
\, . \label{hindf}
\end{equation}
The notation used here reflects the fact that (\ref{hindf}) is a
particular case of the R\'{e}nyi $\alpha$-entropy \cite{cbtw15}.
Combining (\ref{prf25}) with convexity and the decrease of the
function $x\mapsto\!{}-\ln{x}$, we claim the following.

\newtheorem{prp3}[prp1]{Proposition}
\begin{prp3}\label{pan3}
Let $\cmb=\bigl\{\clb_{1},\ldots,\clb_{M}\bigr\}$ be a set of MUBs
in the $d$-dimensional Hilbert space $\hh$. For any state $\bro$, the
average min-entropy obeys
\begin{equation}
\frac{1}{M}\,\sum_{\clb\in\cmb} H_{\infty}(\clb|\bro)
\geq\ln\biggl(\frac{M\sqrt{d}}{\sqrt{d}+\sqrt{M^{2}-M}}\biggr)
\, . \label{bound3}
\end{equation}
\end{prp3}

This new form of the uncertainty relations for MUBs should be compared
with the state-independent relation based on (\ref{indk}) and
lemma 3 of \cite{rastmub}. Namely, we have \cite{rastmub}
\begin{equation}
\frac{1}{M}\,\sum_{\clb\in\cmb} H_{\infty}(\clb|\bro)
\geq\ln\biggl(\frac{\sqrt{M}{\,}d}{d+\sqrt{M}-1}\biggr)
\, . \label{rmub12}
\end{equation}
If we are interested in state-independent lower bounds that hold
for all states, then formula (\ref{bound3}) can overcome
(\ref{rmub12}). In general, the competition between (\ref{bound3}) 
and (\ref{rmub12}) is similar to that between relation
(\ref{bound22}) and (\ref{bound21}). For sufficiently
small $M$, lower bound (\ref{bound3}) is stronger.

\section{Conclusions}\label{sec4}

Using the relative entropy of coherence and the geometric
coherence, we obtained the uncertainty relations for the averaged
quantifiers taken with respect to several MUBs. The uncertainty
relations for the averaged relative entropy of coherence are
state-dependent via the purity and the von Neumann entropy
calculated with the given state. Further, the derived lower bounds 
are explicitly expressed in terms of the number of MUBs and the
dimensionality. The presented bounds differ from the uncertainty
relations for the relative entropy of coherence derived in
\cite{pati16,pzflf16}. The authors of \cite{pati16,pzflf16} used
the entropic uncertainty relations obtained in \cite{lmfan15}. The
method of \cite{lmfan15} leads to the appearance of the entropic bound as a
result of solving a certain maximization problem. For 
MUBs, this general lower bound is not always better than the bounds derived just for such bases. The uncertainty relations
for the averaged geometric coherence of a pure state were
also presented. We examined two lower bounds, neither of
which overcomes the other everywhere. To the best of our
knowledge, the uncertainty relations for the geometric coherence have
not been addressed previously. Finally, we considered novel
state-independent uncertainty relations for several MUBs in terms
of the averaged min-entropy.

\end{document}